\documentclass[twocolumn,pra,superscriptaddress,showpacs]{revtex4}
\pdfoutput=1
\usepackage{graphicx}
\usepackage{amsmath}
\usepackage{amssymb}
\usepackage[ansinew]{inputenc}
\usepackage{bm}\let\mathbf\bm
\newcommand{\bea}{\begin{eqnarray}}
\newcommand{\eea}{\end{eqnarray}}

\begin{document}
\title{Quantum Simulation of the Hubbard Model: The Attractive Route}

\author{A.~F.~ Ho}
\affiliation{Department of Physics, Royal Holloway, University of London, Egham, Surrey
TW20 0EX, United Kingdom }
\author{M.~A. Cazalilla}
\affiliation{Centro de F\'{\i}sica de Materiales (CFM). Centro
Mixto CSIC-UPV/EHU. Edificio Korta, Avenida de Tolosa 72, 20018
San Sebasti\'an, Spain}
\affiliation{Donostia International Physics Center (DIPC), Manuel
de Lardiz\'abal 4, 20018 San Sebasti\'an, Spain.}
\author{T. Giamarchi}
\affiliation{University of Geneva, DPMC-MaNEP, 24 Quai
Ernest-Ansermet CH-1211 Geneva 4, Switzerland}
%\date{\today}

\begin{abstract}
We study the conditions under which, using a canonical
transformation, the phases sought after for the repulsive
Hubbard model, namely a Mott insulator in the paramagnetic and
anti-ferromagnetic phases, and a putative $d$-wave superfluid
can be deduced from observations in an optical lattice loaded
with a spin-imbalanced ultra-cold Fermi gas with attractive
interactions, thus realizing the \emph{attractive} Hubbard
model. We argue that the Mott insulator and antiferromagnetic
phase of the repulsive Hubbard model are easier to
observe in the attractive Hubbard mode as a band insulator of
Cooper pairs and superfluid phase, respectively.
The putative d-wave superfluid phase of the repulsive
 Hubbard model doped away from half-filling is related to a $d$-wave
antiferromagnetic phase for the \emph{attractive} Hubbard
model. We discuss the advantages of this approach to 'quantum
simulate' the Hubbard model in an optical lattice over the
simulation of the doped Hubbard
model in the repulsive regime. We also point out  a number of
technical difficulties of the proposed approach and, in some
cases, suggest possible solutions.
\end{abstract}

\maketitle

\section{Introduction}

Understanding the  phase diagram of the two-dimensional (2D)
single-band  Hubbard model is considered  by many as the 'Holy
Grail' of the theory of strongly correlated systems. In the
most interesting regime, this model describes a system of
spin-$\frac{1}{2}$ (\emph{i.e.} two species of) fermions
hopping on a 2D square lattice with repulsive (on-site)
interactions, and average lattice filling less than   one
fermion per site. This model has been proposed as the minimal
model that explains the observation of $d$-wave
superconductivity with fairly high critical temperature in the
doped cuprate materials\cite{Anderson87-HTc-RVB} (for a review
on doped Mott insulator, see \cite{Lee04-HTc-review}). At half
filling, where only one particle per site is allowed, it is by
now rather well established that the model is a Mott insulator,
which at low temperatures (below a characteristic scale, the
N\'eel temperature, $T_{\rm N\acute{e}el}$) orders
anti-ferromagnetically. Away from half filling, the nature of
the ground state is a subject of heavy debate.  One of the most
challenging open issues is whether the Hubbard model on a 2D
square lattice would support a $d$-wave superconducting phase
at a (relatively) high temperature. The fact that the Hubbard
model can support such an instability has been theoretically
proven in double-chain systems coupled by hopping (known as
two-leg Hubbard ladders)
\cite{dagotto_2ch_review,BalentsFisher96-2legHubbard,schulz_2chains,giamarchi_book_1d}.
However, whether this result extends to the  2D model
consisting of an infinite number of coupled chains is still
extremely
controversial~\cite{Lee04-HTc-review,Demler-Zhang-review,Controversy}.
At present, neither analytical nor numerical  studies are able
to settle the issue.

Due to the spectacular advances in the optical manipulation
of ultra-cold atomic gases,  one very promising route for  studying  the
low temperature phases of the Hubbard model has opened up recently
\cite{Bloch08-correlated-coldatom-review}. Indeed,  ultra-cold Fermi gases
loaded into an optical lattice  can be regarded as almost ideal quantum simulators of the
Hubbard model, where independent control of the hopping amplitude, $t$,
and the on-site interaction energy, $U$, are both experimentally available.
Exploiting this fact, the Mott insulating phase of the Hubbard model
has  recently been demonstrated in a 3D cubic optical lattice (where the
center is at half-filling) by several experimental  groups
\cite{Jordens08-MI-fermions,Schneider08-3D-fermionic-MI}.
Many other  groups are  currently engaged in similar experimental
endeavors~\cite{DARPA08-2DHubb-coldatoms}, with the main focus on
realizing the repulsive Hubbard model on a 2D square lattice away from half-filling,
namely the regime where $d$-wave superfluidity is speculated to exist.
However, one of the main problems that lie ahead in this program has to do with
the currently accessible temperatures for the Fermi gases in optical lattices.
At present, these temperatures (of the order of a few tenths of the Fermi energy of
a non-interacting gas of similar average density) still largely exceed the
N\'eel temperature  ($T_{\rm N\acute{e}el}$), thereby washing out any
 anti-ferromagnetic order in the half-filled system\cite{Ho08-conncet-cluster}.

 However, there are other problems with the present approach to simulate
the repulsive Hubbard model, which seem not to have received so
much attention thus far. One of the most remarkable ones  is the
difficulty of doping away from half-filling a
Fermi gas loaded in an optical  lattice.
In this case, the situation is very different  from doping  in solids,
mainly because of two reasons: i) The existence of an
overall harmonic trapping potential superimposed on the
optical lattice  potential that makes the system inhomogeneous
and tends to favor the maximum site occupancy  (\emph{i.e.} two fermions per site)
near the center  of the trap. As recently demonstrated
experimentally\cite{Jordens08-MI-fermions,Schneider08-3D-fermionic-MI},
for small number of particles this tendency can be
balanced by the onsite repulsion energy, which
yields a Mott insulator near the center surrounded by a 'metallic'
region where the density of holes is
non-uniform.  ii)  Although the  number of available lattice sites can
be controlled with accuracy, the total number of atoms is still hard to measure
accurately and it  is also subject to variations from  shot to shot that are
inherent to the preparation process~\cite{Number_variations}.

 Another source of problems has to do with the need to independently control the
 on-site interaction, $U$, and the hopping amplitude, $t$.  As the N\'eel temperature 
 ($T_{\rm N\acute{e}el}$) (below which the system orders anti-ferromagnetically, and, upon doping,
 the putative $d$-wave superfluid  may appear), scales as
 $T_{\rm N\acute{e}el} \sim t^2/U$, and  thus  rapidly decreases if the
 ratio $t/U$ is made very small by increasing the  optical lattice depth,
it is desirable to have independent control of both $t$ and $U$.
In order  to achieve this, the  $s$-wave scattering length, $a_s$
($U \propto a_s$, roughly speaking) that characterizes the strength and sign of the
 atom-atom interaction, must be tuned towards a Feshbach resonance where
 $a_s \to \pm \infty$.  Since the current interest is in realizing a Hubbard model with repulsive
 interactions,  the side of the Feshbach resonance
where the atom-atom interaction is repulsive ({\it i.e.} $a_s >0$)
must be used  (see, however, Sect.~\ref{sec:realization} for further remarks).
On this side of the resonance, there is a weakly  bound molecular bound
state~\cite{Chin08-Feshbach-review}, with which the  atoms  in the continuum (that is,
in the lowest and highest Bloch bands, when loaded in a lattice)
have a sizable overlap near the resonance.  Thus, at sufficiently low temperatures,
Feshbach molecules form resulting from three atom collisions~\cite{Chin08-Feshbach-review}.
A collision of one of these molecules with a third atom can cause the molecule to make
a transition into a more bound molecular vibrational
state. The released energy is taken away by the colliding atom, which therefore causes undesirable heating
of the system. Also, the presence of these molecules is not accounted for by the single-band
Hubbard model, which is the goal of the quantum simulation.
Furthermore, as the scattering length increases, $U$ also increases and become
of the order of the separation between Bloch bands, thus leading to the break-down
of the single band approximation\cite{DienerHo,Werner06-Hubb-1band-entropy,Ho06}.

   In this article, we propose to explore a different route to simulate the
 Hubbard model in a regime where the onsite interaction is attractive.
 Theoretically, the attractive and repulsive regimes are related by a
 transformation that is well known in the literature of the Hubbard model and
 it is, for completeness,  reviewed in Sect.~\ref{sec:transformation}.
 More recently, in the context of cold atomic gases,  this transformation has been
 used by Moreo and Scalapino~\cite{Scalapino07}, who  pointed out the connection between the
 Fulde-Ferrel-Larkin-Ovchinikov state in the attractive Hubbard model and a state with
 stripes in the repulsive model. These authors also briefly considered
the relationship between the $d$-wave superfluid order
parameter for the repulsive model and a $d$-wave
anti-ferromagnetic order in the  attractive case. For the
one-dimensional Hubbard model (note, however, that there is no
$d$-wave superfluid phase in this case), the transformation was
also used in Ref.~\cite{Luscher08} in an analysis of the noise
correlations of the attractive Hubbard model with spin
imbalance. Moreover, the physics of the attractive Hubbard
model has  also attracted much interest by
itself~\cite{AHubbard1}, especially in recent times  and in
connection with cold atomic gases and the physics of the BEC to
BCS crossover~\cite{AHubbard2}.

In this article, we explore in depth the possibilities
offered by  the attractive model to
understand the physics of the
repulsive Hubbard model. We pay special attention to the   effects of the
trapping potential as well as the peculiarities of the physical
realization of the negative $U$  Hubbard model in optical lattices.
We thus argue that the attractive regime presents a number of advantages for the
quantum simulation of the Hubbard model in an optical lattice.
We also discuss how the (negative-$U$) equivalent phases of the paramagnetic Mott insulator,
the anti-ferromagnetically ordered Mott insulator, and the  putative
$d$-wave superfluid   may be observed.

 We would like to emphasize that one of the main advantages of the
route suggested here is that the equivalence of doping  away from half-filling
the  optical lattice system in the repulsive regime can
be achieved by creating a spin-imbalanced gas. Fermi gases with spin-imbalanced
populations are nowadays routinely created  in the
laboratory~\cite{Ketterle08-Feshbach-BCSBEC-review}, and the magnetization
(which is fixed for the duration of the experiment by the preparation method) can
be controlled to within a few percent accuracy. Other advantages will be
discussed further below.  However, our approach does not solve the problem of how to
achieve lower temperatures for the fermions on the lattice\cite{Ho08-conncet-cluster}.

 The outline of this article is as follows. In Sect.~\ref{sec:transformation} we discuss the
 transformation that formally maps the repulsive  Hubbard model into an attractive one.
 We also define the two attractive Hubbard models we shall be concerned with in this
 paper, together with their corresponding repulsive models.
 We also describe how various physical operators and order parameters are affected
 by the mapping. Some important caveats concerning the realization of the attractive Hubbard model
 are discussed in Sect.~\ref{sec:realization}. In Sect.~\ref{sec:Mott} we discuss the
 equivalent state of the paramagnetic phase  of the Mott
 insulator (as well as possible ways of detecting it), whereas in Sect.~\ref{sec:antiferro}
 we do the same for the equivalent state of the anti-ferromagnetically ordered Mott insulator.
 The effect of doping with  holes, which, as mentioned above, amounts to a spin-imbalanced
 situation in the attractive case,  is analyzed in Sect.~\ref{sec:doping}. Finally, in Sect.~\ref{sec:conclusions}
 we offer the conclusions of the present work as well as mentioning some open problems.

\section{The Hubbard model and the particle-hole transformation}  \label{sec:transformation}

The Hamiltonian of the single-band Hubbard model reads:
\begin{equation}\label{Ham}
 H =  - t \sum_{\langle {\bf i j} \rangle \;\sigma}
c^{\dagger}_{{\bf i} \sigma}  c_{{\bf j} \sigma} + U \sum_{\bf
i} ( n_{{\bf i} \uparrow} - \frac{1}{2})
( n_{{\bf i} \downarrow} - \frac{1}{2} ) + H_{\rm ext},  \;
\end{equation}
where $t$ is the hopping amplitude and $U$ the one-site
interaction. We consider here only the case where the sites
${\bf i}$ of the lattice constitute an hypercubic lattice
(square in two dimensions, and cubic in three dimensions);
$\langle i,j\rangle$ in the  hopping term means that the sum
runs over nearest neighbor sites only. In the above equation,
$n_{{\bf i} \sigma}=c^{\dagger}_{{\bf i} \sigma} c_{{\bf i}
\sigma} $ is the occupancy of spin $\sigma = \uparrow,
\downarrow$ fermions at the $i$-th site. For simplicity and unless
otherwise stated, we will refer in the following to the
two dimensional case, and thus to
a square lattice. All our results are straightforwardly
generalizable to the case of any hypercubic lattice.

We have denoted as $H_{\rm ext}$ all the external fields like
chemical and trap potentials as well as an external Zeeman (magnetic)
field that act upon the system\footnote{We wish to point out for the solid state community reader
that the effect of the physical magnetic field is purely to change the Zeeman energy of the
hyperfine states of the two species of fermions considered here. In particular, since
cold atoms are neutral, there are no orbital effects on the superfluid, such as the Meissner effect
for a superconductor. Henceforth, we shall call the physical magnetic field the Zeeman field, to emphasize
this.}.
Their effects will be discussed below. In the grand canonical ensemble,
\begin{equation}
 H_{\rm ext} = - \mu N + h M
\end{equation}
where  the total number operator $N = \sum_{i,\sigma}
n_{i\sigma}$ and the total magnetization, $M = \sum_{i} \left(
n_{i\uparrow} - n_{i\downarrow} \right)$, and $\mu$
along with $h$ are determined by the condition that averages of $N$ and
$M$ over the grand canonical ensemble yield the experimentally
observed values~\footnote{Note that
here because the interacting term contains a
term proportional to $N$, the definition of $\mu$ is somewhat non-standard.}.
However, cold atomic gases are prepared in
eigenstates of both $N$ and $M$, and therefore the relevant
ensemble is the canonical instead of the grand canonical.
Although it is important to keep this distinction in mind, we
expect that for sufficiently large $N$, the results of both
ensembles  coincide and thus, we shall use the grand canonical
for the calculation of the experimental signatures of the
different phases to be described below in Sect.~\ref{sec:doping}.

The Hamiltonian of Eq.~(\ref{Ham}) has been written in a form
such that for a uniform system the ground state will have
exactly one particle per site, at any temperature, for $U > 0$ and $H_{\rm ext}  =0$.
However,  note that  in real experiments cold-atomic gases are harmonically trapped.
Therefore, the most general form of $H_{\rm ext}$ reads:
\begin{equation} \label{eq:chemmag}
 H_{\rm ext} = \sum_i (\epsilon_i
-\mu)(n_{i\uparrow} + n_{i\downarrow} - 1) - \sum_i  h_i \left(n_{i\uparrow} - n_{i\downarrow} \right),
\end{equation}
where $\epsilon_{\bf i}$ is the shift in the local chemical
potential caused by the trap. We have added an unimportant
constant to the total energy ( $ = \sum_i (\epsilon_i-\mu)$ ),
which will become convenient further below. In current optical
lattice experiments,  we have $\epsilon_{\bf i} = \frac12
\epsilon_0 {\bf i}^2$, but more general forms of the trap may
become available in the future. In the case relevant to
experiments, the Zeeman field $h_i = h$ is uniform (and it is
used to adjust the total magnetization in the grand canonical
ensemble). However, in Eq.~(\ref{eq:chemmag}) we have assumed
it to be site-dependent for further convenience.

We next note that, formally, the sign of the interaction term
(the term $\propto U$  in Eq.~\ref{Ham}) can be changed by
means of the following (particle-hole) transformation on a
bipartite lattice such as the 2D square lattice:
\begin{equation} \label{canon}
\begin{split}
    c_{i \downarrow} & = c_{i_x i_y \downarrow}
    \longleftrightarrow (-1)^{i_x+i_y} \; c_{i \downarrow}^{\dagger}   \\
    c_{i \uparrow} & \longleftrightarrow  c_{i \uparrow}.
\end{split}
\end{equation}
Note that the transformation leaves the operators of the spin
$\uparrow$ fermions unchanged. However, it affects the spin
$\downarrow$ occupation operator: $n_{i \downarrow}
\leftrightarrow 1-n_{i \downarrow}$, and thus the sign of the
interaction term changes ($U \leftrightarrow -U$)  while
hopping term is left invariant (the minus sign in the
right-hand side of Eq.~(\ref{canon}) takes care of this).
However,  it will be important for the discussion that follows
that the transformation exchanges the roles of $h_i$ and
$(\epsilon_i-\mu)$ in Eq.~(\ref{eq:chemmag}). Mathematically,
\begin{equation}
\begin{split}
H_{\rm ext} \to H^{\prime}_{\rm ext} = \sum_{i}
(\epsilon_i-\mu)(n_{i\uparrow} - n_{i\downarrow})  \\
  - \sum_i h_i \left(n_{i\uparrow} + n_{i\downarrow} - 1\right). \label{eq:hamext}
\end{split}
\end{equation}
If we insist in using the point of view of the canonical ensemble, the transformation
(\ref{canon}) implies that $N \to N'= M + {\cal N}$ and $M \to  M' = N  - {\cal N}$,
where $\cal N$ is the total number of lattice sites. Thus, if we consider
an unmagnetized (\emph{i.e.} $M = 0$)
system where $N = (1 + x) {\cal N}$ ($x$ being the doping, where
$x = 0$ corresponds to half-filling in the uniform case), we have that
$N' =  {\cal N}$ and $M' = x {\cal N}$. In words, the doped lattice at $U > 0$ away
from half-filling maps onto a  $U < 0$ system at finite magnetization.
We emphasize that, as discussed below,
the details of the order that the system develops depend not just on
usual factors such as the temperature and strength of the trapping and lattice potential, but are also
constrained by  these globally conserved quantities.

\begin{figure}[tb]
\includegraphics[width= \columnwidth]{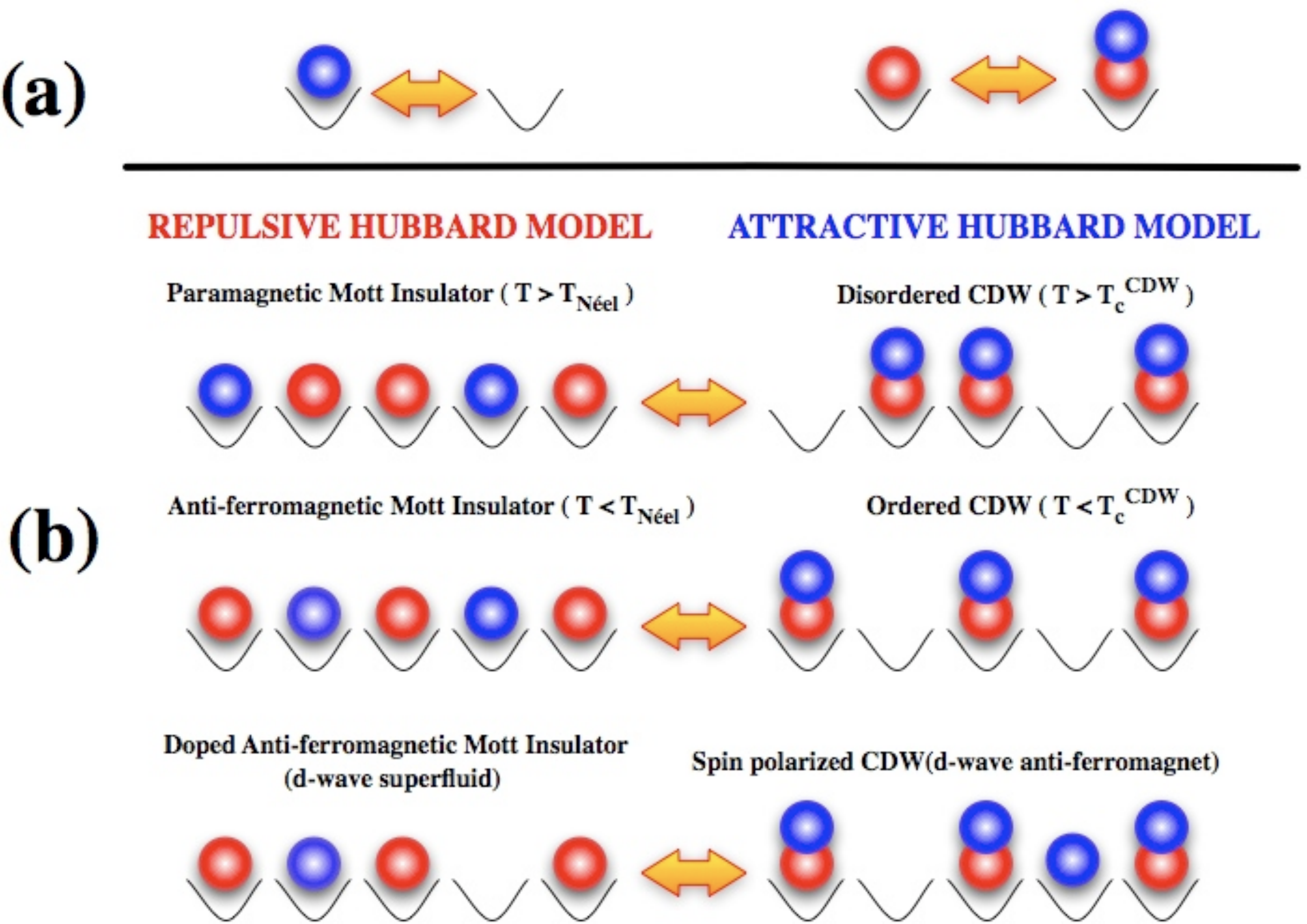}
\caption{(Color online)
Schematic diagram of how the transformation of Eq.~(\ref{canon}) works. In (a) we
sketch how it acts on a single site (\emph{e.g.} red stands for spin up and blue for
spin down), and in (b), the way it transforms different types of states on
a uniform optical lattice.  In addition to the phases depicted above,
an anti-ferromagnetic state at $U > 0$ ordered along the $x$ or
$y$ direction corresponds to a superfluid phase of fermion pairs.
The bottom diagram illustrates one of the main points of this
paper, namely that doping the attractive Hubbard model corresponds to introducing
spin inbalance in the repulsive Hubbard model.}
\label{fig:map}
\end{figure}

It is also convenient to recall that,  in momentum
space, the transformation of Eq.~(\ref{canon}) becomes:
\begin{equation}
    c_{{\bf k} \downarrow} \leftrightarrow c^{\dagger}_{{\bf
    k+Q} \downarrow} \; , \qquad c_{{\bf k} \uparrow}
    \leftrightarrow c_{{\bf k} \uparrow}  \;\; ,
    \label{eq:canon-k}
\end{equation}
where $Q=(\pi/a, \pi/a)$ is the nesting vector with $a$ the lattice spacing.
This expression can be used to obtain the way the different order parameters
and the corresponding phases  transform between the $U > 0$ and the $U < 0$
cases. This is shown in table~\ref{table1}. The way the transformation affects the
different phases expected for the Hubbard model is also illustrated in Fig.~\ref{fig:map}.

Finally, for the sake of clarity, we will spell out in what follows  the two physically
different attractive Hubbard models considered in this article, as well as the repulsive
Hubbard models onto which, via the particle-hole transformation Eq.~\ref{canon}, they are mapped.
The first attractive Hubbard model (from here on called model A1) has
the following Hamiltonian (in the grand canonical ensemble):
\bea \label{eq:A1}
H_{\rm A1} &=& - t \sum_{\langle {\bf i j} \rangle \;\sigma}
c^{\dagger}_{{\bf i} \sigma}  c_{{\bf j} \sigma} - |U| \sum_{\bf
i} ( n_{{\bf i} \uparrow} - \frac{1}{2})
( n_{{\bf i} \downarrow} - \frac{1}{2} ) \nonumber \\
& & + \sum_{{\bf i}} \left(\epsilon_{\bf i} -\mu \right)
\left(  n_{{\bf i} \uparrow} + n_{{\bf i} \downarrow} -1 \right) \nonumber \\
& & - h \sum_{{\bf i}} \left(  n_{{\bf i} \uparrow} - n_{{\bf i} \downarrow} \right).
\eea
This is the Hamiltonian that describes cold atomic systems trapped
in an optical lattice,  with the overall harmonic trapping potential $\epsilon_{\bf i} =
\frac{1}{2} \epsilon_0 i^2$, and a uniform Zeeman field that  can be viewed as a knob to
tune the spin imbalance (See Sect.~\ref{sec:realization}). In this article,
we argue that simulating this Hamiltonian
has a number of advantages over current attempts at simulating
the repulsive Hubbard model in presence of the harmonic trap
(from here on called model R2), whose Hamiltonian reads:
\bea \label{eq:R2}
H_{\rm R2} &=& - t \sum_{\langle {\bf i j} \rangle \;\sigma}
c^{\dagger}_{{\bf i} \sigma}  c_{{\bf j} \sigma} + |U| \sum_{\bf
i} ( n_{{\bf i} \uparrow} - \frac{1}{2})
( n_{{\bf i} \downarrow} - \frac{1}{2} ) \nonumber \\
& & +\sum_{{\bf i}} \left(\epsilon_{\bf i} -\mu \right)
\left(  n_{{\bf i} \uparrow} + n_{{\bf i} \downarrow} -1 \right) .
\eea
The other attractive Hubbard model (from here on called model A2) is
described by  the following Hamiltonian:
\bea \label{eq:A2}
H_{\rm A2} &=& - t \sum_{\langle {\bf i j} \rangle \;\sigma}
c^{\dagger}_{{\bf i} \sigma}  c_{{\bf j} \sigma} - |U| \sum_{\bf
i} ( n_{{\bf i} \uparrow} - \frac{1}{2})
( n_{{\bf i} \downarrow} - \frac{1}{2} ) \nonumber \\
& & +\sum_{{\bf i}} \left(\epsilon_{\bf i} -\mu \right)
\left(  n_{{\bf i} \uparrow} - n_{{\bf i} \downarrow} \right) .
\eea
This is the Hamiltonian that can be obtained from the repulsive model R2 (Eq.~\ref{eq:R2}) via the
particle-hole transformation of Eq.~(\ref{canon}). Model A2 has an inhomogeneous Zeeman field
stemming (via the transformation) from the trapping potential of model R2. In this article,
 model A2 is used to help us understand
\emph{e.g.}  the coexistence of phases in the current experimental regime of the repulsive Hubbard model (R2).
Note that the attractive model A1 that we are advocating does \emph{not} map onto the currently studied
repulsive Hubbard model R2, but instead onto the repulsive Hubbard model
in a inhomogeneous Zeeman field (called model R1 here):
\bea \label{eq:R1}
H_{\rm R1} &=& - t \sum_{\langle {\bf i j} \rangle \;\sigma}
c^{\dagger}_{{\bf i} \sigma}  c_{{\bf j} \sigma} + |U| \sum_{\bf
i} ( n_{{\bf i} \uparrow} - \frac{1}{2})
( n_{{\bf i} \downarrow} - \frac{1}{2} ) \nonumber \\
& & + \sum_{{\bf i}} \left(\epsilon_{\bf i} -\mu \right)
\left(  n_{{\bf i} \uparrow} - n_{{\bf i} \downarrow}  \right) \nonumber \\
& & - h \sum_{{\bf i}} \left(  n_{{\bf i} \uparrow} + n_{{\bf i} \downarrow} -1 \right).
\eea

Thus, to summarize,  using the mathematical  transformation of
Eq.~(\ref{canon}), we can relate the physics of the attractive
Hubbard models to that of the repulsive ones.  In particular
the observation of one particular phase (\emph{e.g.} a $d$-wave
AF phase) at $U < 0$ would directly  imply the existence of the
corresponding phase at $U > 0$ (the putative $d$-wave
superfluid phase). Below we shall see that the realization of
some of these phases in the attractive regime requires
sometimes less stringent conditions than the corresponding ones
in the repulsive regime. Furthermore, as described above, the
exchange of roles of the Zeeman field and the chemical
potential terms effected by the transformation  implies that we
can simulate the doping of the Mott insulator by creating a
system with a finite magnetization (\emph{i.e.} a
spin-imbalanced system). Also, even if the  temperatures that
can be achieved in current experiments do not allow for the
investigation of the low-temperature ordered states, we may
expect that, by the attractive route, some  useful insights can
be gained into  other controversial issues for the high-$T_ c$
community, such as the nature of the normal state of the 2D
repulsive Hubbard model away from half-filling.
\begin{table}[htdp]
\caption{Transformation between phases (and their associated order parameters)
of the Hubbard model for $U>0$ and $U<0$  under the canonical transformation
of Eq.~(\ref{canon}) (cf.  Eq.~\ref{eq:canon-k} in momentum space).
In the expressions below,  $\phi^{(d)}_{\bf k} = \left( \cos k_x - \cos k_y \right)$
is the $d_{x^2 - y^2}$ lattice  form factor.}
\begin{center}
\begin{tabular}{|c|c|}
\hline
$U>0$ & $U<0$  \\ \hline \hline
paramagnetic Mott & disordered lattice with \\
insulator (MI, $n_i = 1$)  & $n_i =0$ or $n_i = 2$  \\ \hline
$s$-wave AF MI (z) & s-CDW insulator \\
$M^{(s) z}_{\bf Q} = \sum_{\bf k}  \sum_{\sigma} \sigma \langle
c^{\dagger}_{{\bf k} \sigma} c_{{\bf k+Q} \sigma} \rangle$  &
$\Delta^{(s) {\rm CDW}}_{\bf Q} = \sum_{\bf k}  \sum_{\sigma} \langle
c^{\dagger}_{{\bf k} \sigma} c_{{\bf k+Q} \sigma} \rangle$ \\\hline
$s$-wave AF MI (x,y) & $s$-wave superfluid     \\
$M^{(s) -}_{\bf Q}=\sum_{\bf k}  \langle c^{\dagger}_{{\bf k+Q} \downarrow} c_{{\bf k} \uparrow} \rangle$
  &  $ \Delta^{(s)} = \sum_{\bf k}   \langle
c_{-{\bf k} \downarrow} c_{{\bf k} \uparrow}\rangle$  \\ \hline
$d$-wave superfluid (?) & $d$-wave AF metal (?) \\
$\Delta^{(d)} = \sum_{\bf k}  \phi^{(d)}_{\bf k} \langle
c_{-{\bf k} \downarrow} c_{{\bf k} \uparrow}\rangle  $ &
$ M^{(d) -}_{\bf Q}=\sum_{\bf k} \phi^{(d)}_{\bf k}  \langle c^{\dagger}_{{\bf k+Q} \downarrow} c_{{\bf k} \uparrow}
\rangle   $
 \\ \hline
\end{tabular}
\end{center}
\label{table1}
\end{table}

\section{Realization of the attractive Hubbard model} \label{sec:realization}

In principle,  since the onsite interaction energy $U$ is \emph{na\"ively} proportional
to the atomic scattering length in free space~\cite{Werner06-Hubb-1band-entropy}, $a_s$,
the $U<0$ regime can be accessed by sweeping
the magnetic field to the side of an inter-species Feshbach
resonance \cite{Chin08-Feshbach-review}
where $a_s < 0$. In the literature of the BCS to BEC
crossover~\cite{BCSBEC,Ketterle08-Feshbach-BCSBEC-review,Chin08-Feshbach-review},
this side is known as the ``BCS side'' of the resonance.

 However, the above point of view entirely neglects the
subtleties of the scattering problem on a lattice potential,
as it turns out that $U$ is not a linear function of $a_s$, in
the general case~\cite{Fedichev04,Wouters06,unpub}.
The details of the dependence
of the  zero momentum scattering amplitude, $f_0(a_s)$ (and $U \propto f_0$),
on the atomic scattering length, $a_s$, are
determined by the dimensionality and other parameters of the
lattice~\cite{Fedichev04,Wouters06,unpub}.
However, all these results share one common feature,
namely the existence of a  particular length  scale, $l_{*}  > 0$,
such that for $a_s = -l_{*}$ the scattering amplitude $f_0$  exhibits a  (geometric)
resonance~\cite{Fedichev04,Wouters06,unpub}: $f_0 \propto a_s /(1+a_s /l_*)$.
 Indeed, the resonance can be approached
either by changing the lattice parameters, or by  changing $a_s$ through a
Feshbach resonance.  We shall focus on the latter case here.
 To realize the attractive Hubbard model,  $U < 0$, we require tuning the scattering
length to the attractive side ($a_s < 0$) but such that $|a_s| < l_{*}$.
As  we approach the Feshbach resonance and $f_0$ diverges to  $-\infty$,
crossing the geometric resonance beyond $l_*$ leads to
the  interaction becoming effectively repulsive, and as a consequence, close
to the geometric resonance, a weakly bound state appears.
The existence of this bound state has been discussed
in the literature~\cite{Fedichev04,Wouters06}, and if the temperature is sufficiently
low (compared to the binding energy of the bound state), many bound states
will be created even if the effective interaction between the atoms in the lattice
is repulsive. This regime is clearly not described by the repulsive Hubbard model,
because it does not take into account the bound states. The situation is similar
(although probably less harmful for the system~\cite{Fedichev04})
to the one encountered as $a_s \to +\infty$,
on the so-called BEC side of the Feshbach resonance.
In this regime, the scattering amplitude corresponds to that of a
repulsive effective interaction, which may  lead us to think that the system
is described by the repulsive Hubbard model, except crucially, for the existence of  
the lattice molecular bound  states. But indeed, as described in  the introduction, the existence of
Feshbach molecules~\cite{Chin08-Feshbach-review} leads to inelastic losses.

To summarize, the attractive regime can be reached by making the scattering
length $a_s$ negative, but not beyond a certain limit where for $a_s = -l_{*}$,
($l_*$ depending on the lattice dimensionality and other
parameters~\cite{Fedichev04,Wouters06,unpub}) the scattering amplitude
has a geometric resonance.

\section{Analog of the Mott insulator in the attractive regime} \label{sec:Mott}

Let us first start by looking at the $U<0$ system with a
balanced population of spin up and down fermions
(Model A1, Eq.~\ref{eq:A1} with $h=0$)
and at temperatures $T$ where
$|U|$ can be made such that $T < |U|$ (however, in this section $T$ is
assumed to be large compared with $t^2/|U|$). Using the
transformation, Eq.~(\ref{canon}),
this corresponds  to a \emph{half-filled} system for $U > 0$
(model R1, Eq.~\ref{eq:R1} with $h=0$). This is the
situation where it is known that  a Mott insulator appears in model R1.
For $U>0$ the existence of the Mott insulator means that the
states with more than one particle per site are strongly
disfavored due to the large on-site repulsion, $U$. The corresponding situation
for $U<0$ is that  the only allowed states for every lattice site are either zero or doubly
occupied states, as shown in Fig.\ref{fig:map}. The existence of
the Mott insulator in the repulsive regime thus corresponds  in the attractive regime
to  having  all fermions form pairs\footnote{By pairs it is meant in this article a many-body bound
state of the same nature as a Cooper pair. These pairs should be distinguished from molecular bound
states arising from poles in the scattering amplitude  of the two-body problem.}.
For sufficiently large $|U|$ these pairs are tightly bound, which means that their
existence can be probed by sending photons to
photo-associate  them into dimers, which
are no longer trapped, and therefore can be detected as a loss
of atoms from the lattice\cite{Ryu05-photoassoc-optlatt}.

\begin{figure}[hb]
%\begin{minipage}[t]{8.0cm}
%\begin{center}
\includegraphics[width=0.9 \columnwidth]{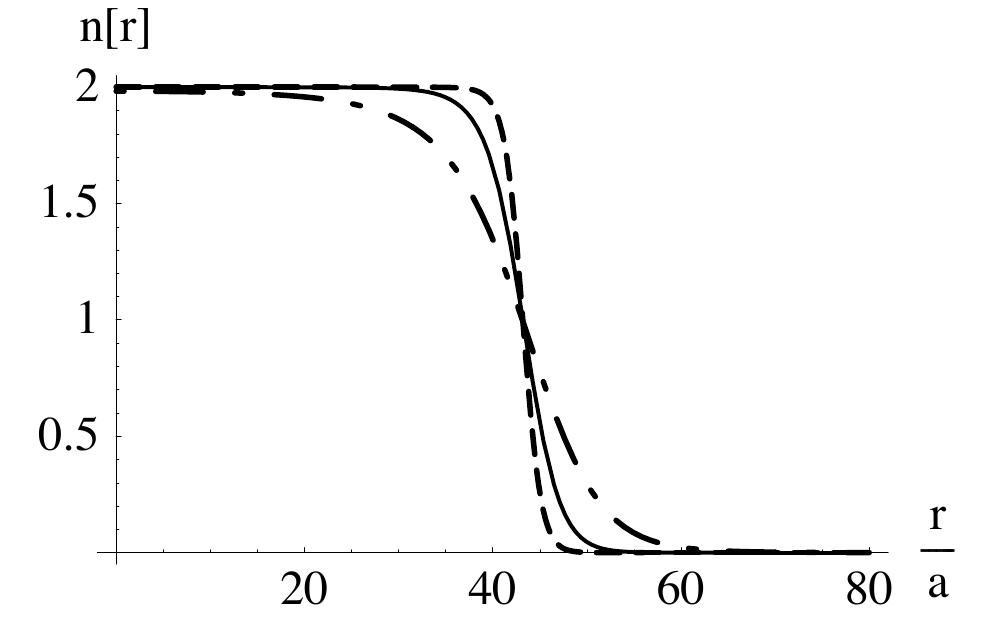}
\caption{Density profile for the attractive Hubbard model in a harmonic trap
(model A1 at $h=0$),
as a function of the radial distance from the trap centre.  Dashed line:
$T/U$=0.05, solid line: $T/U=0.1$, dot-dashed line: $T/U=0.2$. Total number of particles is N=11728. Trapping energy $\epsilon_0/2U = 0.0003$.}
\label{fig:densityprofile-negU}
%\end{center}
%\end{minipage}
\end{figure}

It is worth noticing that this system of pairs exhibits a
pairing gap ($\sim |U|$, for large $|U|$) to spin excitations.
Therefore, a measurement of the single particle properties,
like the single-particle spectral function which is accessible
by photoemission-like  spectroscopy proposed in~\cite{Dao07}
and recently applied to   ultra-cold Fermi gases by the JILA
group~\cite{JinARPES}, should be able to detect the pairing
gap. However, since  in model A1 the harmonic trap only couples
to the atomic density (cf.  Eq.~\ref{eq:A1}), it does not lead
to the breaking of the pairs and, therefore, it  does not take
the system out  of the subspace where $n_i = 0$ or $n_i = 2$
(which, by virtue of Eq.~\ref{canon}  corresponds to
half-filling, \emph{i.e.} $n_i = 1$). Indeed, the role of the
trap is to lift the large energy degeneracy (for $t = 0$) of
the states in this subspace. In other words,  in absence of the
trapping potential in model A1, and since the chemical
potential $\mu$ couples to the \emph{total} particle number,
all states with the same total number of fermions and only
doubly occupied or empty sites are degenerate. The trap  breaks
this degeneracy and  selects as the ground state of model A1
the state where all pairs uniformly occupy the lattice sites at
the trap bottom. In the strong coupling limit, this state can
be regarded as a `band insulator' of the pairs (see
Fig.~\ref{fig:densityprofile-negU}).

 A complementary way of arriving at the same conclusion
relies on the transformation of Eq.~(\ref{canon}). After the
transformation,  model A1 at $h=0$ is mapped to model R1,
that is a repulsive Hubbard model but in a \emph{inhomogeneous}
Zeeman field (note that $h =0$ in this model too, but in this case
it couples to the total density). Although the Zeeman field
affects the magnetic ordering by ferromagnetically polarizing the fermions
at the center of the trap, it does not lead to doubly occupied sites,
and therefore it does not affect the characteristic incompressibility of the
Mott insulator, which depends on the existence of a gap ($\sim U$, for large $U$)
to all density excitations.

It is worth to contrast the situation described in previous paragraphs
with the  one found in current experiments, which are performed in the repulsive regime of the
Hubbard model (model R2, Eq.~\ref{eq:R2}). In such a system, one needs to adjust
the chemical potential $\mu$ (that is, the number of fermions, $N$)
in order to have a half-filled lattice with one fermion per site at the trap bottom.
Otherwise, at too large a  $\mu/U$ (\emph{i.e.} large $N$), the system energetically prefers
to pay the energy cost of  having doubly occupied sites near the bottom of the trap,
rather than accumulating them far from the center, where the trapping energy is very large.
Thus, the lattice at the center of the trap ceases to be in the Mott
insulating phase, becoming a band insulator. On the other hand, for $\mu$ too small
(\emph{i.e.} small $N$) the optical lattice  is not  uniformly occupied by one fermion site.
Testing for the existence of the Mott insulator at $U>0$ thus requires
checking for the absence of doubly occupied sites, which has
been already achieved experimentally by the Z\"urich and Mainz
groups~\cite{Jordens08-MI-fermions,Schneider08-3D-fermionic-MI}.
However, testing the absence of holes (which may appear due to
thermal or quantum fluctuations, especially as $N$ or $U$ decrase)
is a more difficult task. In this regard,
in the attractive regime, only the absence of
singly occupied states needs to be tested. Such measurement,
which implies recording
the spatial   distribution of lattice sites with different
occupations may be accessible through spectroscopic techniques
similar to those employed to observe the `wedding-cake' structure
of the Bosonic Mott insulator~\cite{Folling06_weddingcake,Campbell06_weddingcake}.

\begin{figure}[htb]
%\begin{minipage}[t]{8.0cm}
%\begin{center}
\includegraphics[width=0.9 \columnwidth]{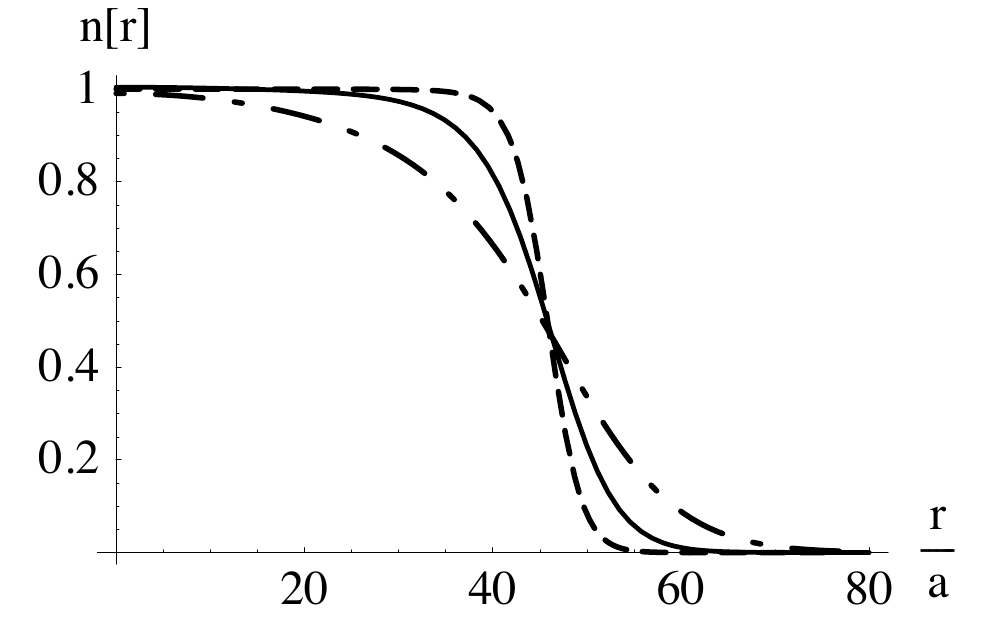}
\caption{Density profile for the repulsive Hubbard model in a harmonic trap (model R2),
as a function of the radial distance from the trap center $r$ in units of $a$, the optical lattice spacing. Dashed line:
$T/U$=0.05, solid line: $T/U=0.1$, dot-dashed line: $T/U=0.2$. Total number of particles is N=6599.
Trapping energy $\epsilon_0/2U = 0.0003$.}
\label{fig:densityprofile-posU}
%\end{center}
%\end{minipage}
\end{figure}

In Figs.~\ref{fig:densityprofile-negU},\ref{fig:densityprofile-posU}
we show illustrative plots (in the large $|U|/t$ limit and for $T < |U|$)
of the  experimentally measurable/measured density profiles
$n(r) = n_{\uparrow}(r) + n_{\downarrow}(r)$, both in the attractive regime
with no spin imbalance (Fig.~\ref{fig:densityprofile-negU} for model A1 at $h=0$) and
in the repulsive regime, the half-filled lattice (Fig.~\ref{fig:densityprofile-posU} for model R2).
\footnote{In the large $U/t$ limit, we can ignore the kinetic energy
and hence, the Hamiltonian is purely local. The density profile is then:
$\langle n_{i \uparrow} + n_{i \downarrow} \rangle =
2 [ \exp\beta\mu_i  +\exp\beta(2\mu_i - s |U|) ]
 / [ 1 + 2\exp\beta\mu_i  + \exp\beta(2\mu_i - s |U|) ]$, where $\mu_i = \mu -\epsilon_i + s|U|/2 $, and
$s= -1$ for the attractive model A1 at $h=0$, and $s= +1$ for the repulsive model R2.
The total number of fermions $N = \sum_i  \langle n_{i \uparrow} + n_{i \downarrow}\rangle $,
which fixes the chemical potential $\mu(T,U)$. We have also found that, at low temperatures, the
density profiles for $U < 0$ can be obtained from a mapping to a free fermion model, details of this
model and its consequences will be reported elsewhere~\cite{unpub}.}
We have used data similar to those in current experiments, \emph{e.g.} the Zurich group
\cite{Jordens08-MI-fermions} (see figure captions).

It is noticeable that  for $U>0$ (model R2), for the given number of particles vs.
trap energy $\epsilon_0 =
m \omega^2 a^2$ ($m$ is the mass of a single fermion and $\omega$ is the trapping frequency),
there is single
occupancy in the central region, which is the Mott insulator region.
At temperatures that are low compared to $U$, but higher than any second order
 (exchange, etc.) processes at the energy scale $\sim t^2 / U$
({\emph i.e.} at currently achievable temperatures), the $\uparrow$ and $\downarrow$
fermions are equally likely to be found on any site of the Mott-insulating region, as there is no
magnetic ordering.  The latter can only emerge as the temperature
is lowered below the N\'eel scale, $T_{Neel} \sim t^2 /U$.  As we move towards
the edge of the sample, the site occupancy  deviates noticeably from one fermion per site,
and a ``metallic'' shell appears. Its width depends on the temperature, as well as $t, U$ and
the trapping energy.

  On the other hand,  for the $U<0$ case of model A1 at $h=0$,   the center of the lattice
 is filled with  fermion pairs resulting from the attractive on-site interaction.
Thus, in model A1 (Fig.~\ref{fig:densityprofile-negU}) the
effect of temperature (in the limit where one can disregard the
second order effects due to the hopping), is to create a finite
density of empty sites near the edge of the 'band insulator' of
pairs. This explains the deviation of the site occupancy  from
$n(r) = 2$ observed in   Fig.~\ref{fig:densityprofile-negU}, as
the distance to the center, $r$, increases. We also notice
that, because the trap is not in competition with the pairing
gap as described above,   the stability of the band insulator
state is not threatened by the trap (contrary to the Mott
insulator at $U > 0$), and thus the fermion numbers in this
regime ($U<0$) can be larger than those at $U > 0$, which is
also illustrated by the sizes used to generate the figures.
However, for model R2, the trap, as described above, tends to
favor  the band insulator in the middle for large enough $N$.
When the total number of fermions $N$ becomes smaller than a
critical value (see next section),  consideration of the
effects of hopping will be required.

\section{Antiferromagnetic order for $U>0$ and its analog in attractive case} \label{sec:antiferro}

As discussed in the previous section,  for large on-site attraction, the
only two possible states  for a single site are empty sites and
doubly occupied. Singly occupied sites are separated from these states by the (large)
energy gap $\sim |U|$. Thus all the fermions are paired and can be regarded as
hard core bosonic entities hopping from site to site with amplitude $\sim t^2/|U|$.
To see this, recall that, in the repulsive case \cite{Auerbach_book}
and ignoring for the moment the
overall harmonic trapping, the low energy effective model for the $U>0$ half-filled
Hubbard model (with equal proportion of the two species) is a
spin-1/2 nearest-neighbor Heisenberg model $H_{\rm eff}(U>0) = J
\sum_{\langle i,j\rangle} {\bf S}_i  \cdot {\bf S}_j,
\;\;J=4t^2/|U|$. This  model transforms for
$U<0$, upon applying (\ref{canon}), into a
half-filled lattice described by the following effective
Hamiltonian in terms of  hard-core bosons  ($b_i = c_{i
\downarrow} c_{i \uparrow}$):
\bea H_{\rm eff}(U<0) = J
\sum_{\langle i,j\rangle}  \left[
(b^{\dagger}_i b_i -\frac{1}{2}) (b^{\dagger}_j b_j -\frac{1}{2}) -b^{\dagger}_i b_j \right],
\label{Heff-boson}
\eea
where $\langle i,j \rangle$ means that the sum runs over nearest
neighbor sites only. For the $U>0$ case, the ground state of the Heisenberg model is
a (s-wave) antiferromagnet (sAF). In the absence of
any terms in the Hamiltonian that  distinguish spin
up and down fermions, that is, for a spin-isotropic Hamiltonian, 
the staggered magnetization  can point in any  of the spin directions, $x,y$,
or $z$.

For the $U<0$ case, the resulting system, Eq.
(\ref{Heff-boson}), is thus a system of  hard-core bosons
(arising from the pairing of two different spin fermions)
hopping on the lattice with the kinetic energy $\sim J$, and
experiencing nearest neighbor repulsion interaction also  $\sim
J$. Several phases can be realized in such a system. The
resulting nearest neighbor repulsion and kinetic energy in
Eq.~(\ref{Heff-boson}) favors an alternating pattern of empty
and doubly occupied states, {\it i.e.}, the checkerboard state
in 2D also known as the "charge density wave" (CDW) state. This
alternating pattern of empty and doubly occupied states for
$U<0$ is the analogue of the antiferromagnetic (along
$z$-direction) state for $U>0$ as shown in Fig.\ref{fig:map}
and Table~\ref{table1}. Another alternative ground state is
simply an s-wave superfluid of these bosonic pairs (sSF). In
fact, the degeneracy for ordering along any direction $x,y,z$
of the sAF for the $U>0$ case maps to a degeneracy between the
CDW state and the SF state for the $U<0$ case, which one can
see formally from the order parameter mapping,  using the
transformation Eq.(\ref{eq:canon-k}): the checkerboard order
parameter $\Delta^{\rm CDW}$ maps to the antiferromagnetic
order parameter in the $z$ spin direction $M^{z}_{{\rm stag}
}$, while the SF order parameter $\Delta^{(s)}, \Delta^{(s)
\dagger}$ maps to the AF order along $-, +$ direction,
$M^{+}_{{\rm stag}} = M^{x}_{{\rm stag}} + i M^{y}_{{\rm
stag}}$ and $M^{-}_{{\rm stag}} = M^{x}_{{\rm stag}} - i
M^{y}_{{\rm stag}}$ \footnote{ Another way to see the same
correspondence is using wavefunctions. A cartoon ground state
for AF ordering in the $x$ direction is of the form:
$|AF_x\rangle \sim \Pi_j [ c^{\dagger}_{j \uparrow} + (-1)^j
c^{\dagger}_{j \downarrow}] |{\rm vac}_c\rangle$, where $|{\rm
vac}_c\rangle$ is the vaccum state for the $c$-fermions.
Defining $d^{\dagger}_{j \uparrow} = c^{\dagger}_{j \uparrow} $
and $d_{j \downarrow} = (-1)^j c^{\dagger}_{j \downarrow} $,
with $|{\rm vac}_c\rangle = \Pi_j d^{\dagger}_{j \downarrow}
|{\rm vac}_d\rangle$, then, $|AF_x\rangle \longrightarrow \Pi_j
[ d^{\dagger}_{j \uparrow} d^{\dagger}_{j \downarrow} + 1 ]
|{\rm vac}_d\rangle$, which is a coherent state of pairs
$b^{\dagger}_j =  d^{\dagger}_{j \uparrow} d^{\dagger}_{j
\downarrow} $.} See Table~\ref{table1}. 

\begin{figure}[tb]
\includegraphics[width= \columnwidth]{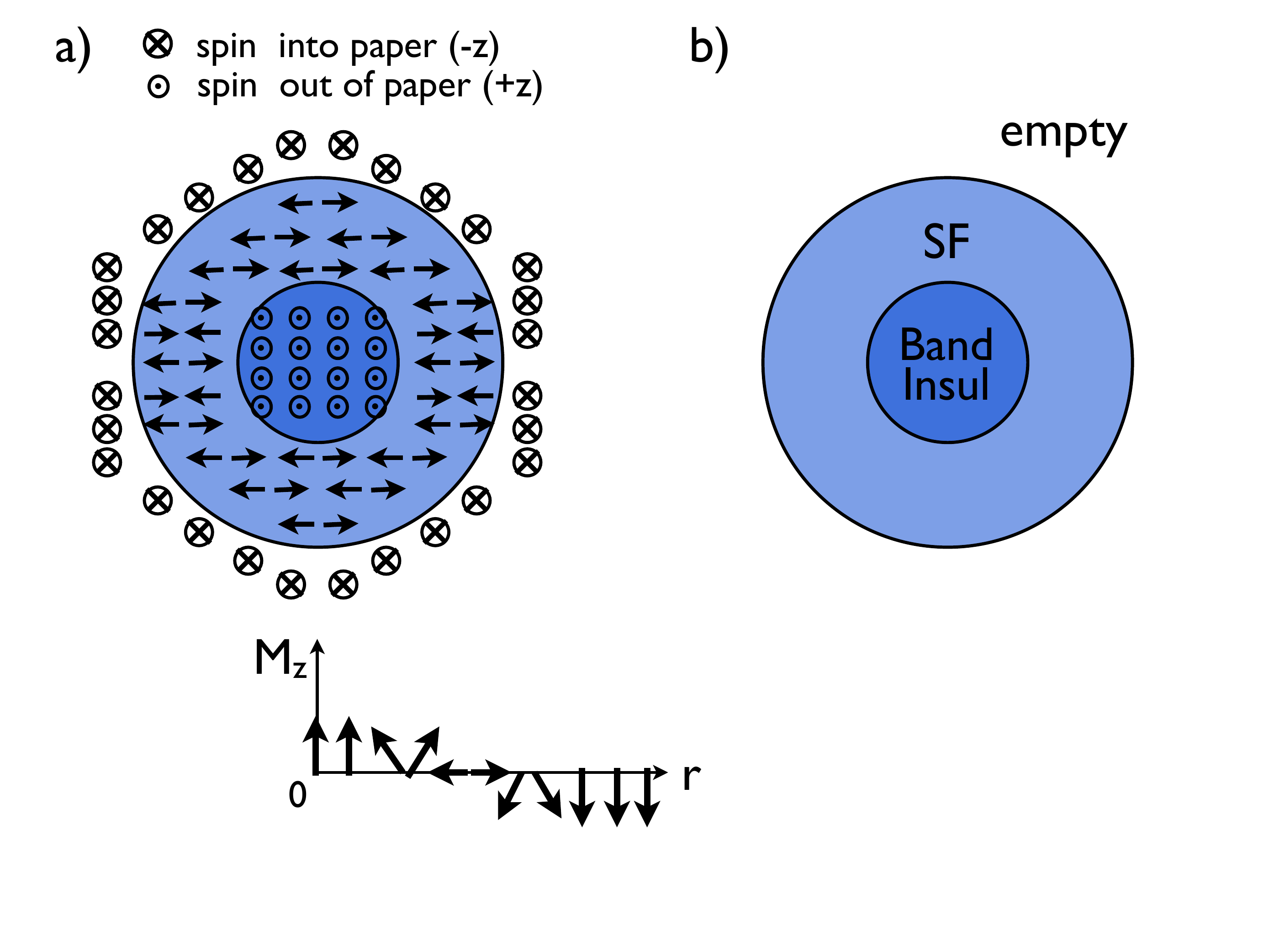}
\caption{(Color online) Schematic diagram of the shell structure of a) the half-filled
repulsive Hubbard model in a site-dependent Zeeman field (model R1), and b) the corresponding
attractive model A1 state configuration in a trap. }
\label{fig:half-fill-trap-config}
\end{figure}

However, the presence of the
harmonic trap on the attractive side leads to interesting
effects. As can be seen from Eq.~(\ref{eq:chemmag}), in model
A1 Eq.~(\ref{eq:A1}), the trap acts as a chemical potential for
the pairs in the $U<0$ case, and thus favors a completely
filled center of pairs (which is just a band insulator in the
center, cf. Fig.~\ref{fig:densityprofile-negU}), rather than
either a CDW or superfluid state. In fact, in the limit where
tunneling between sites is suppressed, the ground state in the
trap potential is this band insulator, and only the kinetic
energy and the nearest neighbor repulsion of order $J$ prevent
this state to occur. Mapping back to the $U>0$ case, one sees
that the trap transforms into a Zeeman field along the $z$
direction to become model R1 Eq.~(\ref{eq:R1}). This Zeeman
field will lift the degeneracy between the various magnetic
states and then polarize \emph{ferromagnetically} the spins
rather than favor an antiferromagnetic order. This competition
is summarized in Fig.~\ref{fig:half-fill-trap-config}.

This effective Zeeman field in model R1 goes from large and
positive in the center of the trap to large and negative in the
periphery. In the center of the trap one would thus have all
the spins polarized up. This phase corresponds via the
transformation, to a pair on each site and thus to the band
insulator of pairs in model A1. Whether the effective field in
the center of the trap is enough to polarize fully the spin
depends on the number of particles in the trap and will be
discussed below. Further from the center, the Zeeman field
decreases and becomes negative. One has thus in a certain
radius a shell where the spins are not fully polarized. In this
region the spins preserve an antiferromagnetic order in the
direction \emph{perpendicular} to the Zeeman field albeit with
a reduced amplitude (\emph{i.e.} the AF order parameters
$M_{Q}^{(s) \pm}$ are non-zero). In other words, in this shell,
the system of model R1  thus possesses antiferromagnetic order
in the $x,y$ direction in spin space. After the transformation
this $x,y$-antiferromagnetic phase for $U>0$ maps for $U<0$ to
an s-wave superfluid phase, as shown in
Fig.~\ref{fig:half-fill-trap-config}. We thus see that looking
for antiferromagnetism in the repulsive Hubbard model amounts
to probing for superfluidity in the $U<0$ one. Then beyond a
certain radius the trap prevents the pair to exist, and the
system is empty (see Fig.~\ref{fig:densityprofile-negU}). In
the $U>0$ system, this corresponds to a Zeeman field that is
large enough to fully polarize to down the system of spins
\footnote{Note that for small $U$ the interplay of the
inhomogeneous magnetic field and the kinetic energy leads to
interesting effects for the intermediate phase, due to the
difference between the radial and angular directions for
the hopping of a particle. These effects will be discussed
elsewhere \cite{unpub}}.

In summary, on the $U<0$ side the trap does not really spoil
the search for the analogue of the antiferromagnetic phase. The
corresponding phase is now a simple s-wave superfluid that can
be probed by similar techniques that have already been  used in
the continuum. The extra phases induced by the trap only
potentially reduce the spatial extent of the superfluid shell,
but since they both corresponds to insulating regions (either
band insulator or empty region), they should not spoil the
observation of the superfluidity. To minimize the central band
insulator region for the attractive model A1 (and hence
maximise the superfluid signal), the chemical potential at the
trap center must not be too strong, so that when tunneling is
turned on, the resulting pair hopping and pair repulsion energy
scale $J$ (see Eq.~\ref{Heff-boson}) can overcome the trapping
potential to delocalize the band insulator. First, in the
strong coupling limit, consider the case where the whole trap
consists of mostly the band insulator. We can estimate the
instability of the band insulator  to this scale $J$: taking a
pair at the edge of the band insulator system ($R = R_2$) to a
site just beyond the edge (by a lattice spacing $a$) costs an
energy $\sim \epsilon_0  (R_2/a)$, and this is to be balanced
against $J= 4 t^2/|U|$. For a large enough system, $\pi
(R_2/a)^2 = N/2$ where $N/2$ is the total number of pairs. Thus
the band insulator state will become unstable when roughly, $
t^2/|U| > \epsilon_0  \sqrt{N/2\pi}$. However, for a full SF
state in the central region of the trap, one needs to move
outward more than just the pairs at the boundary of the band
insulator. A similar estimation indicates that the $\sqrt{N}$
in the above criterion is replaced by linear in $N$. Thus for a
given trap energy $\epsilon_0$, there is an upper limit to
total number of fermions that can be loaded into the trap.
 For example, using data similar to those in the $^{40}{\rm K}$ experiments of the
Zurich group, at an optical lattice depth of 5 recoil energy for
a laser wavelength of 825 nm, a mean trapping frequency of 80 Hz,
and an (independently tuned) ratio $|U|/t = 8$, the total
number of fermions has to be smaller than $\sim 500$ to have a pure superfluid core.
Remembering that this is only a rough estimate,
this is nevertheless a rather small number to achieve under current conditions\cite{Moritz-personal}.
This critical number can be  larger if we allow for a central region of
band insulator in addition to a  superfluid shell.  Depending on the
sentivity to detect the superfluid shell,
and thus the number of atoms that one needs to have in the superfluid, the
number of fermions can be larger than the above estimate. 
 The situation is better for
the lighter atom $^6{\rm Li}$. Using the same numbers as above, but with a laser
wavelength of 1064 nm, the upper critical number becomes $\sim 7000$, which should be
feasible in current experiments.

As for experimental signature, the observation of the coherence
peak in the momentum distribution should signify the onset of
the BEC of pairs, and superfluidity can be proven when vortices
are observed when rotating the trap and optical lattice
(although to date, vortices have been seen in a rotating bosonic BEC
superfluid in an optical lattice\cite{Tung06-rotate-optlatt-BEC-cortex-pin}). These
are simpler probes than say, using noise correlation\cite{Foelling,Rom06,Greiner05}  
to deduce the broken translation symmetry of the AF state in the $U>0$
case\cite{Altman04}.

\section{Effect of doping} \label{sec:doping}

So far, we have shown that the spin-balanced population for $U<0$
already presents several advantages to tackle the Mott and AF
physics. But one of its main advantages is the possibility to
effectively ``dope'' the $U>0$ system by looking at spin imbalanced
$U<0$ systems. This then allows to settle experimentally the
still controversial issue of the presence or absence of the
d-wave superfluid (dSF) in the repulsive Hubbard model doped
away from half-filling.  As mentioned in the Introduction, this
doping may be difficult to do directly in the $U>0$ system due
to the presence of the overall harmonic trap. Via the
transformation, the $U>0$ model away from half-filling maps to
the $U<0$ model with an effective Zeeman field. In the
context of cold atom experiments, this corresponds to a (fixed)
imbalance of spin-up versus spin-down fermions, which can
readily be achieved to an accuracy of a few percent
currently\cite{K04} (see \cite{Ketterle08-Feshbach-BCSBEC-review} for a review).
%{\it This is a major advantage of studying experimentally the $U<0$ system.}

We now examine some of the observables in that case, and in
particular what would be the consequences of the existence of a
$d$-wave superfluid phase in the repulsive Hubbard model for
the $U<0$ phase. We perform this analysis for the homogeneous
system and will discuss the possible effects of the trap at the
end of this section.

\subsection{Transformation of the operators}

Under the  transformation, Eq.~(\ref{eq:canon-k}), the
superfluid order parameter $\Delta^{(\alpha) \dagger}$ (the
label $\alpha = s,d$ indicates the $s$-wave or $d$-wave
symmetry of the order parameter) maps to the commensurate
antiferromagnetic  order parameter $M^{(\alpha) +}_{{\rm stag}
}$,
\bea \label{eq:op-mapping}
\Delta^{(\alpha) \dagger} = \sum_{\bf k}  \phi^{(\alpha)}_{\bf k}
\langle c^{\dagger}_{{\bf k} \uparrow} c^{\dagger}_{-{\bf k} \downarrow} \rangle \nonumber\\
\longleftrightarrow
\sum_{\bf k} \phi^{(\alpha)}_{\bf k}  \langle c^{\dagger}_{{\bf k} \uparrow} c_{{\bf k+Q} \downarrow}\rangle
=M^{(\alpha) +}_{\bf Q}   \nonumber\\
\Delta^{(\alpha)} = \sum_{\bf k}  \phi^{(\alpha)}_{\bf k} \langle
c_{-{\bf k} \downarrow} c_{{\bf k} \uparrow}\rangle  \nonumber\\
\longleftrightarrow
\sum_{\bf k} \phi^{(\alpha)}_{\bf k}  \langle c^{\dagger}_{{\bf k+Q} \downarrow} c_{{\bf k} \uparrow}
\rangle =M^{(\alpha) -}_{\bf Q}\nonumber
\eea
where previously, $\phi^{(s)}_{\bf k}=1$ is the $s$-wave form factor and
 $\phi^{(d)}_{\bf k} = \left( \cos k_x - \cos k_y \right)$ is the $d_{x^2 - y^2}$ form factor.

Thus, {\it if there is a regime of dSC in the $U>0$ Hubbard model, then correspondingly, there
is a regime of dAF in the $U<0$ model}.  This is the analogue of the well-known mapping
at precisely half-filling of the Hubbard model between the ground states of sSF or sCDW at $U<0$ and the sAF at $U>0$ (see Fig.\ref{fig:map} and
Table~\ref{table1}).

\subsection{Momentum distribution and noise correlation: a mean
field calculation}

In this section, we briefly outline a simple calculation to illustrate experimental signatures
of dAF  states that may exist in the $U<0$ system.
We need to emphasize from the outset that since there are no microscopic analytical
calculation of the dSF nor the dAF  states in the 2D Hubbard model, we will instead
use a toy mean field model that does give rise to such states, in order to calculate some
expected responses such as the momentum distribution and noise correlation\cite{Foelling,Rom06,Greiner05} .
While such a mean field model misses out correlations and quantum fluctuations,
the objective here is to demonstrate that \emph{ the symmetry of the CDW or SF order parameters
has very definite signatures in noise correlation experiments}\cite{Altman04}.
Thus, we simply {\it assume} that the
$U<0$ Hubbard model  acquires a  mean field (MF) form, which  in momentum space becomes:
\bea \label{HMF2}
H^{(\alpha)}_{MF} &=& \sum_{\bf k} \left[
 \left( \epsilon_{{\bf k}\sigma} \right)c_{{\bf k}\sigma}^{\dagger} c_{{\bf k}\sigma}
- h \sum_{\sigma} \sigma c^{\dagger}_{{\bf k}\sigma} c_{{\bf k}\sigma} \right]\\
&  +& \sum_{\bf k} g_Q \; \phi^{(\alpha)}_{\bf k} \left(  c^{\dagger}_{{\bf k} \uparrow} c_{{\bf k+Q}\;\downarrow} M^{(\alpha) -}_{\bf Q}+
M^{(\alpha) +}_{\bf Q} c^{\dagger}_{{\bf k+Q} \;\downarrow} c_{{\bf k}\uparrow} \right) . \nonumber
\end{eqnarray} 
where we have introduced (by hand) the mean field order parameter
Eq. (\ref{eq:op-mapping}), and as before, $\alpha$ labels the $s$ or $d_{x^2 -y^2}$ order parameter.
As usual, via a global gauge transformation, the
order parameter can be chosen to be real:
$M^{(\alpha) -}_{\bf Q} = M^{(\alpha) +}_{\bf Q}=M_{\bf Q} $.
Our main interest is in the phases of the $U>0$ Hubbard mode without Zeeman field
away from half-filling. This then corresponds to the $U<0$ mean field model above (Eq. \ref{HMF2}) at
$\mu = 0$ and finite $h$.
In principle, the nesting wavevector ${\bf Q}$ should be a variational parameter to be determined from the
particular band structure etc. However, the following mean field theory only makes sense if
$ {\bf Q} = (\pi/a,\pi/a)$ is
a commensurate wavevector: the interaction above couples ${\bf k}$ with ${\bf k +Q}$, and this
in turn couples to ${\bf k +2 Q}$ which is the same as ${\bf k}$ only for commensurate $ {\bf Q}$.

This MF Hamiltonian is diagonalized by a Bogoliubov rotation:
\bea
\alpha^{\dagger}_{{\bf k}} & = & \cos (\theta_{\bf k}) \;c_{{\bf k}\uparrow}^{\dagger}
+\sin (\theta_{\bf k}) \;c_{{\bf k+Q}\;\downarrow}^{\dagger} \nonumber \\
\beta^{\dagger}_{{\bf k}} & = & -\sin (\theta_{\bf k}) \;c_{{\bf k}\uparrow}^{\dagger}
+\cos (\theta_{\bf k}) \;c_{{\bf k+Q}\;\downarrow}^{\dagger} \; .
\eea Then, provided
\bea
\cos (\theta_{\bf k}) &=& \left[\frac{1}{2}\left( 1 +
\frac{\epsilon_{{\bf k}\uparrow} -\epsilon_{{\bf k+Q}\;\downarrow}-2 h}{2 \;\Omega_{{\bf k} }}
\right)\right]^{1/2} \nonumber\\
\sin (\theta_{\bf k}) &=& \left[\frac{1}{2}\left( 1 -
\frac{\epsilon_{{\bf k}\uparrow} -\epsilon_{{\bf k+Q}\;\downarrow}-2 h}{2 \;\Omega_{{\bf k} }}
\right)\right]^{1/2} \nonumber\\
\Omega_{{\bf k} } &=& \left[ \left(g_Q\;M_{\bf Q}\;\phi_{\bf k}\right)^2
+\left(\frac{\epsilon_{{\bf k}\uparrow} -\epsilon_{{\bf k+Q}\;\downarrow}-2h}{2}\right)^2 \right]^{1/2} \;,
\eea the MF Hamiltonian becomes
\bea
H_{MF} &=& \sum_{\bf k} \left(
E^{\alpha}_{{\bf k} } \alpha^{\dagger}_{{\bf k}} \alpha_{{\bf k}}
+ E^{\beta}_{{\bf k} }\beta^{\dagger}_{{\bf k}} \beta_{{\bf k}} \right)
\eea with the ``magnetic band'' energies:
\bea
E^{\alpha}_{{\bf k} } &=& \frac{\epsilon_{{\bf k}\uparrow} +\epsilon_{{\bf k+Q}\;\downarrow}}{2}
+ \Omega_{{\bf k} } \nonumber \\
E^{\beta}_{{\bf k} } &=& \frac{\epsilon_{{\bf k}\uparrow} +\epsilon_{{\bf k+Q}\;\downarrow}}{2}
- \Omega_{{\bf k} } \nonumber \; .
\eea The ground state is then made up by filling these bands up to the respective Fermi surface
$k^F_{\alpha \sigma} $ with the occupation number
$n^{\alpha}_{{\bf k} } = \langle AF_{\bf Q}|
\alpha^{\dagger}_{{\bf k} } \alpha_{{\bf k} } |AF_{\bf Q}\rangle
= \Theta (k^F_{\alpha } - k)$ (and similarly for the $\beta$ band):
\bea \label{gsWF2}
|AF_{\bf Q}\rangle = \prod_{\sigma \bf k} \;n^{\alpha}_{{\bf k} } \alpha^{\dagger}_{{\bf k} }
\;n^{\beta}_{{\bf k} } \beta^{\dagger}_{{\bf k} } |0\rangle \;.
\eea Note that this wavefunction does not have definite number $N_{\uparrow}$ and $N_{\downarrow}$
for each species: this is a consequence of the quasi-particles $\alpha$ and $\beta$ carrying indefinite
spin, an analogue of the textbook number non-conserving Bogoliubov quasi-particles or the BCS wavefunction
or the BCS mean field Hamiltonian.
In turn, here for the AF, we have indefinite spin (but definite charge) quasi-particles because we
have assumed the mean field decoupling to be  in the $S_+$-axis in spin space. It is straightforward
to show (at least when $h=0$) that  the same results can be gotten for a spin-conserving mean field wavefunction (eg. when the mean field decoupling is in the $z$ spin axis).

On the square lattice, since $\epsilon_{{\bf k}\sigma} +\epsilon_{{\bf k+Q} \sigma}
= - 2 \mu_{\sigma}$, taking the same chemical potential and bare dispersion for the two
spin species,  $E^{\alpha , \beta}_{{\bf k} } = -\mu \pm \Omega_{{\bf k} }$.
Hence for the $s$-wave case  there is always a band-gap of size $\geq 2 g_Q M_{\bf Q}$, and
the minimum gap occurs at $\epsilon_{{\bf k}\uparrow} -\epsilon_{{\bf k+Q} \downarrow} = 2 h$.
Thus the  half-filled
lattice (1 fermion per site) with $N_{\uparrow}=N_{\downarrow}$ has the $\beta$-band completely filled
up and an empty $\alpha$-band: this is a magnetic insulator since adding one more $\uparrow$ fermion
has $c_{{\bf k}\uparrow}^{\dagger} \rightarrow \cos (\theta_{\bf k}) \;\alpha^{\dagger}_{{\bf k}}$
creating an $\alpha$-particle, but this costs at least the band-gap energy (and similarly for
adding a $\downarrow$ spin particle). For the $d_{x^2-y^2}$-wave case,
since the form factor $\phi_{\bf k} = \cos k_x - \cos k_y$ has nodes at $k_x = \pm k_y$, the band gap
also vanishes at these positions, giving rise to a pseudo-gap only. At less (more) than half-filling,
the ``magnetic Fermi surface'' is within the $\beta$- ($\alpha$-) band and there is no energy gap to
adding an extra fermion: the system is ``metallic''. We should be careful about the meaning of
$E^{\alpha, \beta}_{{\bf k} }$: this is {\bf not} the physical excitation energy (unlike in the
BCS MF theory): in particular, the spin-wave (Goldstone mode) spectrum is missing.

%For a given magnetic field $h$, the polarization
%$P \equiv (N_{\uparrow}-N_{\downarrow})/(N_{\uparrow}+N_{\downarrow})$ is:
%\bea
%P = \frac{\sum_{\bf k} \langle c^{\dagger}_{{\bf k} \uparrow} c_{{\bf k} \uparrow} \rangle
%- \sum_{\bf k} \langle c^{\dagger}_{{\bf k+Q} \downarrow} c_{{\bf k+Q} \downarrow} \rangle}{
%\sum_{\bf k} \langle c^{\dagger}_{{\bf k} \uparrow} c_{{\bf k} \uparrow} \rangle
%+ \sum_{\bf k} \langle c^{\dagger}_{{\bf k+Q} \downarrow} c_{{\bf k+Q} \downarrow} \rangle}
%= \frac{\sum_{\bf k} \frac{\xi_{\bf k} -h}{\Omega_{\bf k}} (n^{\alpha}_{{\bf k} }-n^{\beta}_{{\bf k} })}{\sum_{\bf k}
%(n^{\alpha}_{{\bf k} }+n^{\beta}_{{\bf k} })} ,
%\eea where $\xi_{\bf k}= (\epsilon_{{\bf k}\uparrow} -\epsilon_{{\bf k+Q}\;\downarrow})/2$.

%We plot in Fig.\ref{polar} $P$ vs. $h$ for the dAF.

Using the ground state Eq. (\ref{gsWF2}), the $T=0$ momentum distribution for each spin component is:
\bea \label{nk2}
\langle c_{{\bf k}\uparrow}^{\dagger} c_{{\bf k}\uparrow} \rangle &=& \cos^2 (\theta_{\bf k})
\; n^{\alpha}_{{\bf k} }  + \sin^2 (\theta_{\bf k}) \; n^{\beta}_{{\bf k} } \; \\
\langle c_{{\bf k+Q}\downarrow}^{\dagger} c_{{\bf k+Q}\downarrow} \rangle &=& \sin^2 (\theta_{\bf k})
\; n^{\alpha}_{{\bf k} }  + \cos^2 (\theta_{\bf k}) \; n^{\beta}_{{\bf k} } \; \label{nk2d}
\eea In experiments, assuming we can image
perpendicular to the plane of the lattice, this result
has to be convolved with the Wannier function $w({\bf k})$ (in momentum space) for the optical lattice:
$\langle n^{\sigma}_{{\bf k} =m {\bf R}/\tau} \rangle \propto
|w({\bf k})|^2 \langle c_{{\bf k}\sigma}^{\dagger} c_{{\bf k}\sigma} \rangle$. (When there are more than
1 plane of the lattice, we also need to integrate over the planes.)

\begin{figure}[tb]
%\begin{minipage}[t]{8.0cm}
%\begin{center}
\includegraphics[width=0.9 \columnwidth]{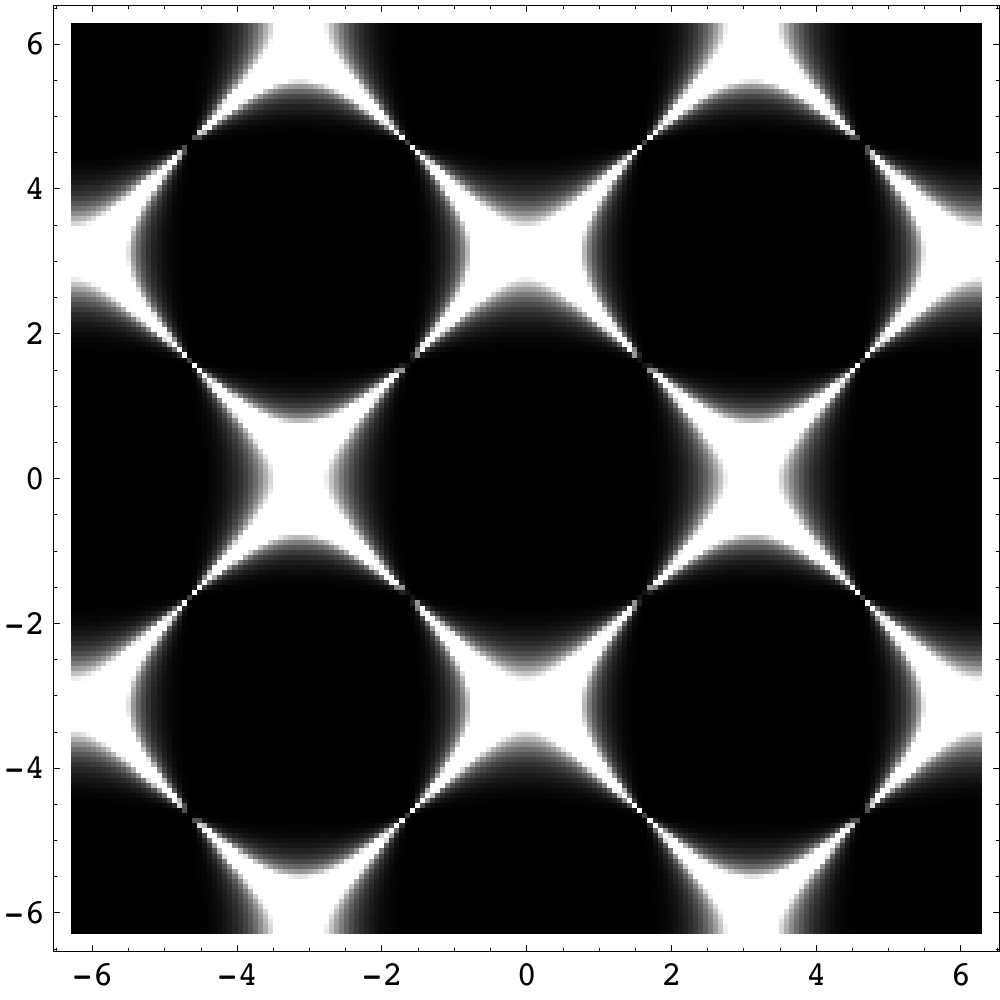}
\caption{Density (grey-scale) plot of the noise correlation for the 2D $d_{x^2-y^2}$-wave AF against $k_x$ and $k_y$
when ${\bf k'}= {\bf k+Q}$. Parameters used: $g_Q M_{\bf Q} = 0.10$, $\mu= 0$, $h=0.23$.}
\label{Cnoise-dAF0-h}
%\end{center}
%\end{minipage}
\end{figure}

The $T=0$ noise correlation is proportional to the connected correlation function:
\bea \label{noise-AF}
G_{\sigma \sigma'}({\bf k,k'}) &=& \langle c_{{\bf k}\sigma}^{\dagger} c_{{\bf k}\sigma}
c_{{\bf k'}\sigma'}^{\dagger} c_{{\bf k'}\sigma'} \rangle -
\langle c_{{\bf k}\sigma}^{\dagger} c_{{\bf k}\sigma} \rangle
\langle c_{{\bf k'}\sigma'}^{\dagger} c_{{\bf k'}\sigma'} \rangle \nonumber \\
 & = & - \delta_{\sigma, -\sigma'} \delta_{{\bf k'}, {\bf k+Q}} \; \sin^2 \theta_{\bf k}
\cos^2 \theta_{\bf k} \left(n^{\beta}_{{\bf k} }
(1- n^{\alpha}_{{\bf k} } )\right) \nonumber \\
&=& \frac{-\delta_{\sigma \;, -\sigma'} \;\delta_{{\bf k'}, {\bf k+Q}}
\left(g_Q\;M_{\bf Q}\;\phi_{\bf k}\right)^2}{4 \left[\left(g_Q\;M_{\bf Q}\;\phi_{\bf k}\right)^2 +
\left(\frac{\epsilon_{{\bf k}\uparrow} -\epsilon_{{\bf k+Q}\;\downarrow}-2h}{2}\right)^2\right]}
\nonumber \\
& & \times \left(n^{\beta}_{{\bf k} } (1- n^{\alpha}_{{\bf k} } )\right)
\eea
(Again, for the experimentally measured noise correlation,
this result has to be multiplied by $|w({\bf k})|^2 |w({\bf k'})|^2$.)
We plot in Fig. \ref{Cnoise-dAF0-h} the noise correlation for the 2D $d_{x^2-y^2}$-wave AF, at $\mu=0$. Noise correlation should clearly distinguish between $s$-wave and $d_{x^2-y^2}$ AF, thanks to the appearance of nodes in the $\pm (\pi/a, \pi/a) $ directions.

%We measure all energies with the units  $ t =1$.
%With finite $h=0.23$, corresponding to polarization $P \approx 0.111 $ (cf. Fig.\ref{polar}),
%the Fermi surfaces are split between the $\uparrow$ and $\downarrow$ components.
%

%\begin{figure}[t]
%\begin{minipage}[t]{8.0cm}
%\begin{center}
%\includegraphics[width=0.9 \columnwidth]{noise-2D-sCDWpipi-mu0A.eps}
%\caption{Density plot of the noise correlation for the 2D $s$-wave CDW against $k_x$ and $k_y$
%when ${\bf k'}= {\bf k+Q}$. Parameters used: $g_Q \Delta_Q = 0.10$,  $\mu= 0$, $h=0$.}
%\label{noise-sCDW}
%\end{center}
%\end{minipage}
%\end{figure}

\subsection{Experimental observation and complications}

We thus see that the negative $U$ side offers the great
advantage to access directly the doped regime, without having
to suffer directly from the presence of the trap. We have
presented in the previous section some possible experimental
signatures that a d-wave superfluid phase would give when
suitably transformed to the negative side. Of course, even if
the situation is potentially improved by the transformation to
the negative $U$ side, probing the doped phase is a
considerable challenge. Some of the limitations are obvious.
The most immediate one is the effect of the temperature.
Indeed, already for the antiferromagnetic phase, the
temperature is in competition with an energy of order $J \simeq
4 t^2/U$ the kinetic energy of the pairs. When looking at the
doping effects one has to face even smaller energies. Lowering
the temperature is thus a must.

The second limitation is again the trap. Although the mapping
to the negative side avoids the direct effect of the trap on
the doped holes, the trap still has a potentially  indirect effect.
Indeed as we saw in the section on antiferromagnetism, the trap
will act, for the $U>0$ side (model R1), as an effective Zeeman field,
and lead to two shell regions, with fully polarized spins up or
down (Fig.~\ref{fig:half-fill-trap-config}). When holes are
introduced into the system of model R1, they will thus have the possibility
to go in one of these two polarized regions or in the
antiferromagnetic region, which corresponds to the shell where
the effective Zeeman field is not strong enough to fully
polarize the system. Where the holes  go will depend on how
much kinetic and interaction energy they can gain in the three
different phases since they are not sensitive directly to the
presence of the Zeeman field. A naive calculation neglecting
the presence of the energy scale $J$, leads directly to the
holes going in the interface between the fully polarized up and
fully polarized down regions. In the $U<0$ language, {\it i.e.} for model A1, this
corresponds to the excess of one spin species going to the edge
of the band insulator region. This seems to suggest, that when
the energy scale $J$ is put back in the problem, the holes will
indeed go into the antiferromagnetic region, and thus can lead
potentially to the d-wave superconducting phase there. This is
however a delicate question since the kinetic energy of a hole
in a ferromagnetic environment is in principle higher due to
the absence of frustration of the antiferromagnetic order upon
hole motion. This important question thus fully deserves more
analytical and numerical study. It remains however academic
until serious progress on the temperature issue have been made.

\section{Summary, conclusion} \label{sec:conclusions}

In this article, we have explored in depth the possibilities
offered by quantum simulating the attractive model (A1) in cold
atoms in optical lattices,  to understand the physics of the
repulsive Hubbard model, via a well known canonical
particle-hole transformation. We have argued that there are
certain advantages in doing experiments in the attractive
regime.

For the undoped case the attractive side replaces the Mott
phase and the antiferromagnetic phase by a phase composed only
of pairs (Mott insulator for $U>0$) and that undergoes a
superfluid transition (antiferromagnet for $U>0$). The trap
which exists in any realistic experiment does not really affect
the observation of these two phases since it can only add a
core of band insulator at the center, thereby not spoiling the
observation of the superfluid. The attractive side also offers
the advantage of only having to test for the pairing for the
observation of the band insulator that must be simpler than
testing for the absence of doubly occupied and empty sites on
the repulsive side.

Another key advantage of the attractive side is the relative
ease in controlling the spin population imbalance. Via the
canonical transformation, this corresponds to doping away from
half filling for the repulsive side, which is hard to achieve
because of the presence of the harmonic trapping potential that
moves the holes away from the central region of the trap. This
doping in the repulsive regime is needed to quantum simulate
and explore the possibility of a d-wave superfluid, the
fundamental question of great relevance to the cuprate high
temperature superconductors. This question can be answered
instead in the attractive regime by exploring the presence or
absence of a d-wave antiferromagnet. We indicate in this paper
several ways to probe for the existence of such a phase. We
have also pointed out a number of technical difficulties of the
proposed approach and, in some cases, suggest possible
solutions.

\acknowledgements 
 TG would like to thank M. Zwierlein for interesting discussions.
We would like to thank H. Moritz for bringing to our attention
an inaccurate estimate in Section V, and for useful correspondence
on the current experimental feasibility.
MAC thanks M. Ueda for his kind hospitality during his visit to 
the University of Tokyo, and the Ueda Macroscopic Quantum Control Project 
of  JST for partial financial support for this visit,
during which parts of this work were done. This work was
supported in part by the Swiss NSF under MaNEP and Division II.
AFH acknowledges financial support from EPSRC(UK) through grant
EP/D070082/1, MAC gratefully acknowledges financial support of
the  Spanish  MEC through grant No. FIS2007-066711-C02-02 and
CSIC through grant No. PIE 200760/007.

\end{document}